\documentclass[5p,twocolumn,authoryear]{elsarticle}
\usepackage[latin1]{inputenc}
\usepackage[english]{babel} 
\usepackage[]{graphicx}
\usepackage[mathcal]{euscript}
\usepackage{amsmath}
\usepackage{amssymb}
\usepackage{amsthm}
\usepackage{natbib}
\graphicspath{{./}{fig/}}
\usepackage{epsfig}
\newcommand{\ba}{\begin{eqnarray}}
\newcommand{\ea}{\end{eqnarray}}
\newcommand{\avg}[1]{\langle #1 \rangle}

\newcommand\e{\emph}
\begin{document}
\title{The shade avoidance syndrome: a non-markovian stochastic growth model}
\author{Andrea Veglio}
\ead{andrea.veglio@unito.it}
\address{Department of Oncological Sciences and Division of 
Vascular Biology, Institute for Cancer Research and Treatment,
University of Torino, Str. Prov. 142 Km 3.95., 10060 Candiolo, Italy.}
\address{National Inter-University Consortium for the Physical Sciences of Matter (CNISM), Corso Duca degli Abruzzi 24, 10129 Torino, Italy.}
\begin{abstract}
Plants at high population density compete for light, showing a series of physiological responses known as the shade avoidance syndrome.
These responses are controlled by the synthesis of the hormone auxin, which is regulated by two signals, an environmental one and an internal one. 
Considering that the auxin signal induces plant growth after a time lag, this work shows that plant growth can be modelled in terms of an energy-like function extremization, provided that the Markov property is not applied.   
The simulated height distributions are bimodal and right skewed, as in real community of plants. In the case of isolated plants, theoretical growth dynamics and speed correctly fit {\it Arabidopsis thaliana} experimental data reported in literature. 
Moreover, the growth dynamics of this model is shown to be consistent with the biomass production function of an independent model. These results suggest that memory effects play a non-negligible role in plant growth processes.   
\end{abstract}
\maketitle

\section{Introduction}
Ecologists have been concerned with competition in plants since the beginning of the last century. \citet{Clements1905} published the 
first formal definition of competition in plants, defining it as ``the relation between plants occupying the same area and dependent upon the same supply of physical factors'', and later \citep{Clements1929} he considered that competition begins ``when the immediate supply of a \e{single factor necessary} falls below the combined demand of the individual plant''.  

Since the 50s a number of authors have systematically studied the effect of competition in plant systems. For example, \citet{Donald1951} investigated in a seminal work the intra-specific light competition in pastures plants.
More recently, \citet{Pacala1996} report experimental data providing little evidence of competition for nitrogen and water, and suggesting that the most important factor determining competition in sufficiently manured and watered plants is light. Accordingly, the present work will assume no stress due to insufficiency of nutrients or water (for models on dryland vegetation, see \citet{Kletter2009} and ref.s therein) and will study light competition in plants grown at high population density, i.e. experiencing density stress. In particular, it will be concerned with intra-specific (inter-plant) light competition.
The series of physiological responses demonstrated by plants experiencing density stress is collectively known as the \e{shade avoidance syndrome} and comprises effects such as stem elongation at expense of leaf and storage organ expansion, inhibition of branching, and acceleration of flowering \citep{Tao2008}. 

There are two main `fingerprints' characterizing the height distribution of a community of plants under density stress:

{\it i.} bimodality: bigger individuals grow more than smaller ones, thus become bigger and bigger, whilst smaller individuals grow much less \citep{Ford1981}.

{\it ii.} hierarchy of exploitation (right skew): there is a large number of plants slightly smaller than the mean, and a small number of plants much greater than the mean, \e{i.e.} the distribution is right (positively) skewed \citep{Harper1967}.

The present work is based on the consideration that growth responses in plants at high population density are controlled by the synthesis of several phytohormones, the most important being auxin \citep{Friml2008,Teale2006}.
Auxin induces plant growth after some delay~\citep{auxin}, and recent experimental results \citep{Tao2008,stepanova2008} show that the auxin pathway is mainly modulated by two signals, an environmental one and an internal one (or, using the vocabulary of \citet{Guedon2007},  ontogenetic). 
The environmental signal consists in the decrease in the red to far-red ratio of incoming light. In fact, plants grown at high population density perceive both absorption of red light by canopy leaves and reflection of far-red light from neighbouring plants \citep{Tao2008}. In other words, a plant `recognizes' to be surrounded by other shading neighbours and responds producing auxin which induces growth after some time lag. 
On the other hand, the internal signal is mediated by the hormone ethylene \citep{stepanova2008}, which stimulates the production of auxin even in isolated plants.

A countless number of models have been developed to study both inter- and intra-specific competition among plants. 
Reviewing the different spatial community dynamics models, \citet{Bolker2003} group them in three different frameworks, namely:

{\it a.} Interacting particle systems (IPS), which are stochastic, markovian \citep{Neuhauser2001}, continuous time models with discrete individuals located in the cells of a regular lattice (see, for example, \citet{Diggle1976,Durrett1994, Hendry1996}). 

{\it b.} Stochastic point processes, which track discrete individuals that compete locally, but assume that they occupy a single point in continuous space (see, for example, \citet{Gandhi1998, Bolker1999, Dieckmann2000}). 

{\it c.} Patch models, which track group of individuals, allowing (noncontiguous patch models) or not (metapopulation or patch-occupancy models) multiple individuals and multiple species per patch (see, for example, \citet{Levins1971, Gilad2004,Gilad2007}).     

Strengths and weaknesses of each of the above frameworks have been exhaustively analyzed by \citet{Bolker2003}. 

In this paper, I propose a model that studies the old problem of the light competition in terms of the old method of the energy-like function extremization, but does not belong to any of the previous frameworks. In fact, even sharing some features with other IPS models, the model presented here is non-markovian. 

Indeed, markovianity is a very nice property in stochastic systems. It assures that in a system the future state depends only on the present state and is independent of past states \citep{Kam07}. In other words, the system has no memory and this considerably simplifies calculations. Unfortunately, auxin production induces plant growth after a time lag, and at the proper time scale this memory effect forbids modelling the system by exploiting the Markov property. 

I will show that in the present model the non-markovianity is necessary:

$\bullet$\ to reproduce plant height distributions that are bimodal; 

$\bullet$\ to correctly mimic experimental growth dynamics in isolated plants;

$\bullet$\ to be consistent with the biomass production model by~\citet{Cournede2008}.

I will consider the total amount of auxin present in a plant, but not its spatial distribution in plant's body, i.e. the well characterized phenomenon of the auxin gradient (see \citet{tanaka2006} and ref.s therein).
For recent models on the auxin gradient the reader is referred to \citet{joensson2006,reuille2006,smith2006,Newell2008}. 

\section{The model}\label{model}
As mentioned in the introduction, auxin is the main regulator of plant growth. For plants growing at high population density, auxin synthesis is regulated by:

1) an environmental signal due to the absorption of red light by canopy leaves and reflection of far-red light from neighboring plants \citep{Tao2008};

2) an internal signal, which makes plants growing also when they do not experience density stress \citep{stepanova2008}. 

Consider a community of plants placed on a 2-dimensional $N\times N$ square lattice. The lattice has $N^2$ nodes, each corresponding to a plant labelled as $(i,j)$, with $i,j=1,\dots, N$. Each $(i,j)$-plant is associated to its own amount of auxin $s_{ij}(t)$ at time $t$. Each plant shades, i.e. interact with, its neighbours by which conversely it is shaded. I will consider only the interaction between first-nearest neighbouring plants. Indeed, trials with second-nearest neighbour interactions showed that the results do not change substantially.   

The auxin amount of a plant is converted into biomass after a delay \citep{auxin} that I will call $\alpha$: if a certain auxin amount is produced by a plant at time $t$, then at time $t+\alpha$ the plant shoots to grow proportionally to the produced quantity of auxin (in this paper I will take into account only vertical shooting, i.e. I will assume the height of each plant to be proportional to its weight.)
Accordingly, I will define the \emph{shade avoidance potential} $a_{ij}(t)$ as the function representing the predisposition at time $t$ of the $(i,j)$-plant to avoid shade by shooting at time $t+\alpha$. Its definition needs some steps. 

For simplicity, first consider an ideal subsystem of two plants, say the $(i,j)$- and the $(i+1,j)$-plant, subjected only to the environmental signal {\it 1)}.

The predisposition at time $t$ of the $(i,j)$-plant to shoot at time $t+\alpha$ depends first of all on its own auxin amount $s_{ij}(t)$. In fact, the greater $s_{ij}(t)$ is, the more the $(i,j)$-plant has the possibility to grow at time $t+\alpha$.  

Moreover, the $(i,j)$-plant perceives an environmental signal from the $(i+1,j)$-plant, that absorbs red light and reflects far-red light. 
This signal ``warns'' the $(i,j)$-plant of the presence of its $(i+1,j)$ neighbour. The greater the $(i+1,j)$ auxin amount $s_{i+1j}(t)$ is at time $t$, the more $(i+1,j)$ will shade the $(i,j)$-plant at time $t+\alpha$. Because the $(i,j)$-plant ``wants'' to avoid the $(i+1,j)$ shade, its predisposition at time $t$ to shoot at time $t+\alpha$ has to depend also on the auxin amount $s_{i+1j}(t)$ of its neighbour. In fact, the greater $s_{i+1j}(t)$ is, the more the $(i,j)$-plant has to grow to avoid the shade of the $(i+1,j)$-plant at time $t+\alpha$. 

Therefore, for a system of two plants subjected only to the environmental signal, the predisposition at time $t$ of the $(i,j)$-plant to shoot at time $t+\alpha$, i.e. the shade avoidance potential $a_{i,j}(t)$, reads 
\ba
a_{ij}(t)=J\ s_{ij}(t)\cdot s_{i+1j}(t).
\ea
$J$ is a non-negative constant that expresses how strong plants shade each other or, in other words, how strong they have to compete for light. 
Because in a generic system of $N^2$ elements the $(i,j)$-plant has more than one first-nearest neighbour (namely, two if $(i,j)$ is at the corner of the lattice, three if it is at the border, four if inside), the $(i,j)$-plant shades and is shaded by all its first nearest neighbours. Therefore, all the contributions from the first-nearest neighbour have to be taken into account in the shade avoidance potential, i.e. 
\ba\label{com}
a_{ij}(t)= J\ s_{ij}(t)\ \sum_{n.n.}s_{ij}(t), 
\ea
where   
$\sum_{n.n.}s_{ij}(t)\equiv s_{i-1j}(t)+s_{i+1j}(t)+s_{ij-1}(t)+s_{ij+1}(t)$ is the sum over the first nearest neighbour of the $(i,j)$-plant.  

Eq.~(\ref{com}) takes into account only the environmental signal {\it 1)} perceived by the $(i,j)$-plant; if it had no neighbours, its shade avoidance potential would be zero, i.e. it would have no chance to produce biomass. Since isolated plants grow, the internal signal {\it 2)} can be taken into account by adding a term proportional to the auxin amount in the plant, $h s_{ij}(t)$, where $h$ is a non-negative constant expressing the strength of the signal, i.e. how high the plant would grow if isolated. Therefore, the complete function expressing the \emph{shade avoidance potential} of the generic $(i,j)$-plant reads
\ba\label{a}
a_{ij}(t)= h\ s_{ij}(t) + J\ s_{ij}(t) \sum_{n.n.}s_{ij}(t).
\ea
Notice that the greater the shade avoidance potential is at time $t$, the more the $(i,j)$-plant will be able to shoot at time $t+\alpha$, i.e. the more it will be able to intercept light and avoid shade. 

Studying the system in the mean field approximation, it is worth considering the \emph{global} shade avoidance potential $A(t)\equiv\sum^N_{i=1}\sum^N_{j=1} a_{ij}(t)$.  
Because all the plants of the community contribute with their mutual shading, $A(t)$ is a global quantity that represents the predisposition of the whole system to avoid shade by shooting after the time lag or, in other words, the {\it mean} shade avoidance potential up to a $N^{-2}$ factor. $A(t)$ reads
\ba\label{hamcl}
A(t)= \sum_{ij}\bigg(s_{ij}(t)\big[h+J \sum_{n.n.}s_{ij}(t)\big]\bigg),
\ea 
where the sum $\sum_{ij}\equiv\sum^N_{i=1}\sum^N_{j=1}$ runs over all the $N^2$ plants. In passing, notice that Eq.~(\ref{hamcl}) is formally analogous to the Ising hamiltonian\footnote{Actually, it is the opposite. In fact, here the global shade avoidance potential will be maximized, while in the Ising model the hamiltonian is minimized.} \citep{Huang1987}.  

Accordingly to what discussed above, it is reasonable to assume that on average plants try to maximize the shade avoidance potential: the greater is the global potential at time $t$, the maximal will be the light exposure of the community at time $t+\alpha$. 

\begin{figure}
\centering
\includegraphics[scale=0.45]{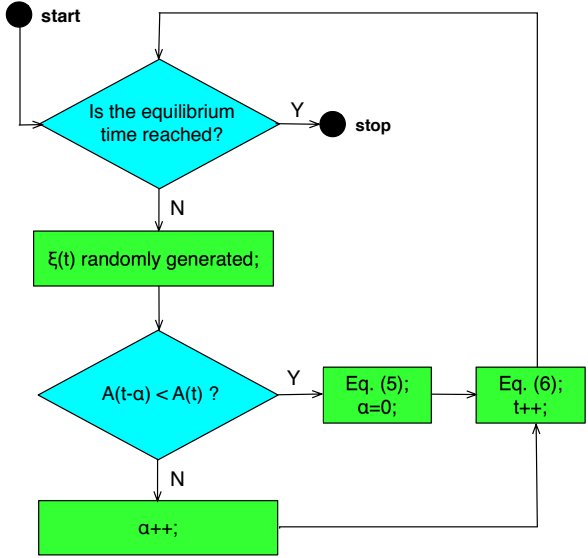}
\caption{Flowchart of the system dynamics.}
\label{flow}
\end{figure}

Therefore, at each time step $t$ the auxin amount of a randomly selected $(i,j)$-plant may be increased only if the global shade avoidance potential is maximized, as follows: 
\ba\label{spin}
s_{ij}(t+1)&=&s_{ij}(t)\\\nonumber&+& \rho\ \xi(t)\ e^{-b t}\ \Theta \big[A(t)-A(t-\alpha)\big]. 
\ea
$\xi(t)\in[-1,1]$ is a randomly drawn number with flat distribution, $\rho>0$ the maximum updating amplitude for the selected auxin amount, $1/b>0$ the equilibrium time\footnote{In this work, for equilibrium I mean the condition when the increments $\Delta s_{ij}$ become negligible. This happens at time $t\sim 1/b$.}. $\Theta[\cdot]$ is the Heaviside step function, defined as $\Theta[x]=0 \mbox{ if } x\le0 \mbox{ and } \Theta[x]=1 \mbox{ if } x>0$. 
The exponential term $e^{-b t}$ has been chosen to let the growing process dramatically slow down once reached the equilibrium time $1/b$.  
The positive discrete variable $\alpha$ represents the time lag introduced above and is calculated according to the flowchart in Fig.~\ref{flow}.
Because at the very beginning of a plant life (dormant seed) the auxin amount is negligible, it is reasonable to choose as initial condition $s_{ij}(0)=0$. 

Eq.~(\ref{spin}) states that, at each time step, the selected auxin amount $s_{ij}$ can be updated by a randomly drawn quantity (which gets smaller and smaller as time goes by because of the exponential term) only if such an update implies the maximization ($\Theta [A(t)-A(t-\alpha)]$) of the shade avoidance potential~(\ref{hamcl}), whose dynamics is defined as: 
\ba\label{ham}
A(t+1)&=&A(t)\\\nonumber&+&\rho\ \xi(t)\ e^{-b t}\ \bigg(h+J\sum_{n.n.}s_{ij}(t)\bigg),
\ea
with $A(0)=0$.\\
$\xi(t)$ is the \e{same} random number selected in (\ref{spin}). 
The sum $\sum_{n.n.}$ runs over the nearest neighbours of the selected $(i,j)$-plant as above.

A mortality rate has been introduced in the following way: if at certain time
the $(i,j)$-plant has reached an auxin amount smaller than a threshold $d$, $s_{ij}$ is set to zero and no longer updated during the simulation. This corresponds to the death of the $(i,j)$-plant. For the sake of clarity, the mortality process has not been shown in the flowchart in Fig.~\ref{flow}.

Now, consider the function $q_{ij}(t)$, that shall represent the height of the $(i,j)$-plant at time $t$. As discussed before, the auxin amount of a plant is converted into biomass after a time lag $\alpha$, and the biomass of a plant is assumed to be proportional to its height, $q_{ij}(t)$. Therefore, setting the proportionality constant equal to one, before the equilibrium is reached the plant height at time $t$ corresponds to the auxin amount at time $t-\alpha$. Then, at the equilibrium, the system does not evolve appreciably and, therefore, the plant height at time $t$ corresponds to the auxin amount at the same time. In formulae, 
\ba\label{qq}
q_{ij}(t)&=& s_{ij}(t-\alpha)\qquad\mbox{for }t< 1/b,\\
q_{ij}(t)&=& s_{ij}(t)\ \ \ \ \ \qquad\mbox{for }t\gtrsim 1/b.
\label{qqq}
\ea 

\section{Numerical results}
I performed numerical simulations of the dynamics \mbox{(\ref{spin}-\ref{ham})} using the following parameter values: $N=10$, $\rho=.2$, $b= 10^{-4}$, $J=10^2$, $h=1$. The mortality threshold has been set to $d=.3$, which corresponds to a rate of mortality of 70\%, accordingly to experiments by \citet{Ford1975}. 
\begin{figure}
\centering
\begin{tabular}{c}
\includegraphics[scale=0.4]{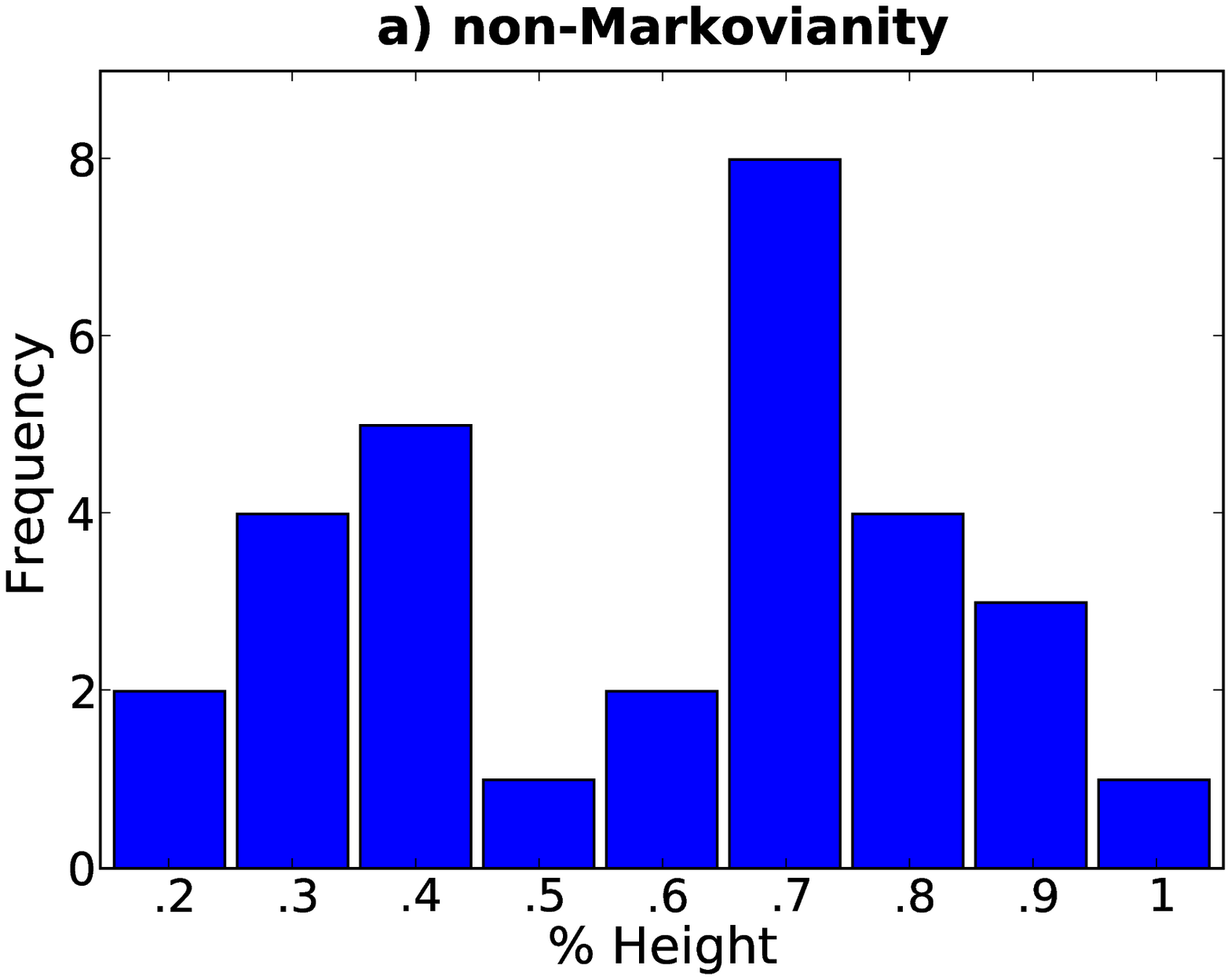}\\
\includegraphics[scale=0.4]{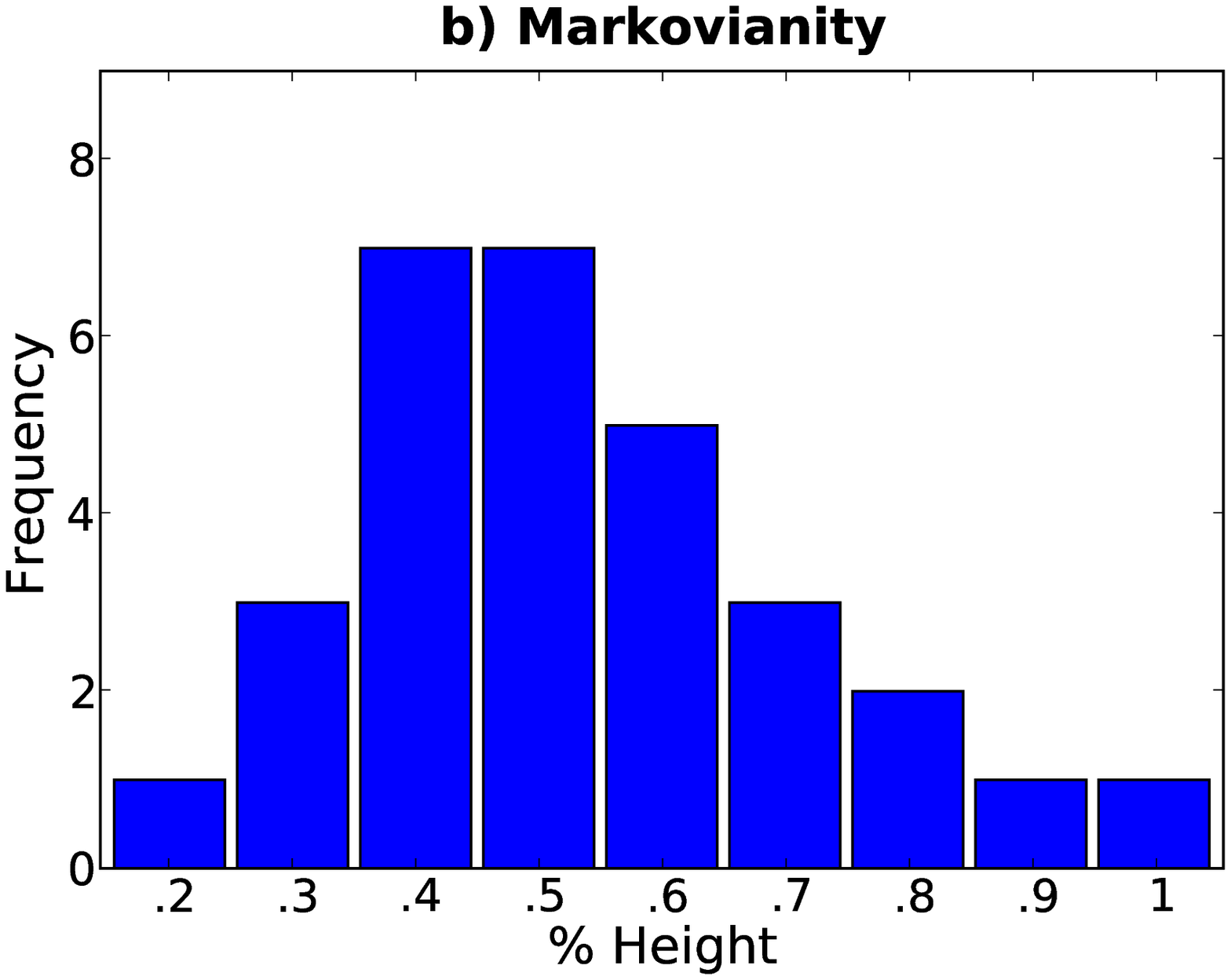}\\
\end{tabular}
\caption{Height distribution at the equilibrium, $t\gtrsim 1/b$, for an initial community of $N^2=100$ plants. The rate of mortality have been set to 70\%, accordingly to experiments by \citet{Ford1975}. The heights are reported in percentage relatively to the maximal height in each sample. a) Simulation of the system according to the model: the Markov property is relaxed. The distribution is bimodal and right-skewed, as in the experiments by \citet{Ford1975}. b) Simulation of the system setting the time lag $\alpha=0$, i.e. forcing the system to be Markovian. The distribution looses bimodality.}
\label{gramma}
\end{figure}

Fig.~\ref{gramma} shows two height distributions at the equilibrium, i.e. at $t\gtrsim 1/b$, for initial communities of $N^2=100$ plants. Each histogram represents the height distribution of the $dN^2=30$ survived plants. 

In Fig.~\ref{gramma}a the simulation has been performed according to the model presented in the previous section: the system is non-markovian. The distribution results to be bimodal and right-skewed (see experiments by \citet{Ford1975}).
Moreover, plants at the lattice border are on average (see Appendix~\ref{app1}) shorter than central ones. 

On the contrary, in Fig.~\ref{gramma}b, the time lag $\alpha$ has been set to be equal to zero, \emph{i.e.} the system has been forced to be markovian. The height distribution looses bimodality, showing that in this model framework the non-markovianity is a fundamental feature to interpret data of plants competing for light. 

The numerical simulations show that $\alpha$ varies with time. In particular, $\alpha$ can be consistently fitted by the simple saturating function 
\ba\label{alpha}
\alpha(t)=\frac{m\ t}{K+t},
\ea  
where $m$ is the plateau value reached by $\alpha$ and $K\equiv1/(2b)$.

\section{Isolated plants}
The model presented here is intended to mimic the behaviour of plants at high population density. Indeed, it may be extended to isolated plants.

In Appendix~\ref{app0}, I derived the expression for the growth (height) dynamics $q(t)$ and the growth speed $\dot{q}(t)$ in isolated plants. They read
\ba\label{heightSec}
{q}(t)&=&{\rho\ \Omega}\ b^{-1}\ (1-e^{-bf(t)}),\\
\label{solgammaSec}
\dot{q}(t)&=&\rho\ \Omega\ \Big(1-\frac{m\ K}{(K+t)^2}\Big)\ e^{-bf(t)},
\ea 
where $\Omega$ is defined as in~(\ref{lambda}) and  
\ba\label{ff}
f(t)=t-\alpha(t)=\frac{K-m+t}{K/t+1}
\ea

This allows to directly check the model consistency with data available in literature.

\citet{Jouve1998} report experimental data for {\it Arabidopsis thaliana} floral stem elongation over time. 
They study circadian rhythms and report experimental data for Arabidopsis first inflorescence internode length and for first inflorescence extension rate as functions of time. 
The first inflorescence internode length can be assumed to be proportional to the plant growth dynamics $q(t)$, and the first inflorescence extension rate, up to a constant, to the growth speed $\dot{q}(t)$. 

\begin{figure}
\centering
\includegraphics[scale=0.45]{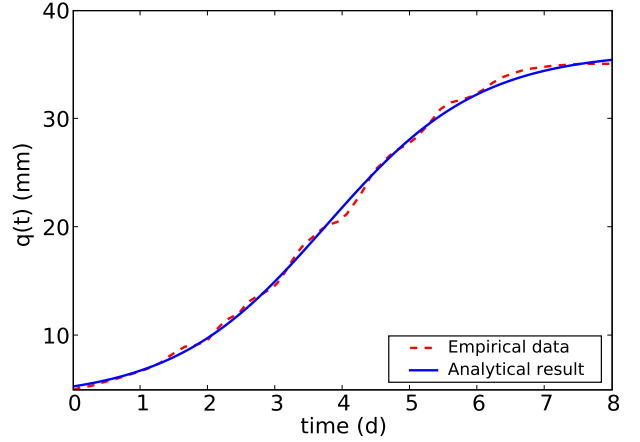}
\caption{\emph{Arabidopsis thaliana} growth dynamics. 
Eq.~(\ref{solgammaSec}) (continuous line) and experimental data from Fig.~1, dataset~\emph{A} (12 hours of dark every 12 hours of light), in \citet{Jouve1998} (dashed line).}
\label{dataq}
\end{figure}

\begin{figure}
\centering
\includegraphics[scale=0.45]{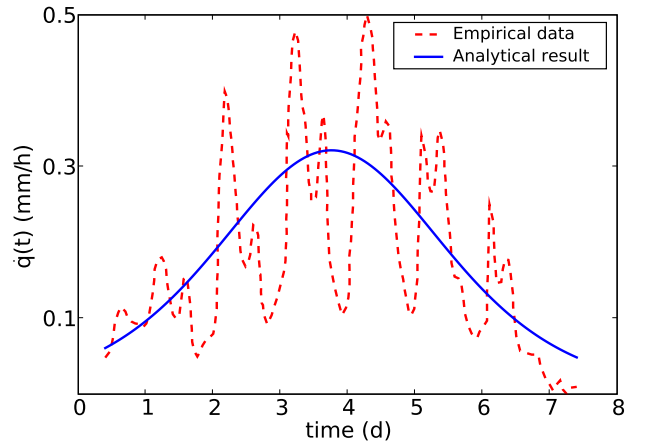}
\caption{\emph{Arabidopsis thaliana} growth speed. 
Eq.~(\ref{heightSec}) (continous line) and experimental data from Fig.~2A in~\citet{Jouve1998} (dashed line).}
\label{datas}
\end{figure}

Fig.~\ref{dataq} shows $q(t)$~(\ref{heightSec}) fitted on Arabidopsis first inflorescence internode length data and Fig.~\ref{datas} shows $\dot{q}(t)$~(\ref{solgammaSec}) fitted on Arabidopsis first inflorescence extension rate data. 

The agreement are satisfactory in both cases; fluctuations in first inflorescence extension rate data are due to the circadian rhythm, which is not taken into account by this model and which can be easily averaged out.  

Interestingly, \citet{Cournede2008} propose the following empirical formula for the biomass production $Q(t)$, 
\ba\label{found}
Q(t)=\beta\ \lambda(t)\ (1-e^{-\gamma S(t)}),
\ea
where $\beta$ and $\gamma$ are characteristic parameters, $S(t)$ the leaf surface area of the plants, and $\lambda(t)$ a function of environmental conditions related to the evapotranspiration that I here assume to be constant.

Considering again the Arabidopsis case, experimental data from 
Fig.3a in \citet{Cookson2007} show that, in the time window of interest (0-8 days), Arabidopsis leaf surface area over time can be consistently fitted (reduced $\chi^2\sim1$) by $a f(t)$, being $a$ a proportionality constant. 

This suggests that the growth dynamics $q(t)$ obtained here and the biomass production $Q(t)$ proposed by \citet{Cournede2008} are consistent. While parameter identification between the two models needs further investigation, it is worth observing that $f(t)$ can correctly reproduce experimental data for leaf surface area over time only in the case $m\neq 0$. 
In other terms, it is necessary the system to keep memory of the time lag $\alpha$ even in the case of isolated plants.

\section{Sensitivity analysis}
Focusing back to the general model ($J\ge0$, $N\ge1$), it is interesting to analyze its parameter sensitivity.

I have performed a stochastic analysis of the model (Appendix~\ref{app1}), whose main result is, according to~(\ref{s}) and~(\ref{l}), 
\ba\label{fof}
\langle s_{ij}\rangle_{st}\simeq\mbox{function($\rho,b,N$)}.
\ea
This means that on average the auxin amount in a generic $(i,j)$-plant does not directly depend on $J$ and $h$. Numerical simulations confirmed this result. Speculating, this suggests that plant height does not directly depend either on the R:FR ratio or on the hormone ethylene density. This observation is consistent with the fact that plant sensitivity both to light \citep{Maloof2001} and to hormone ethylene \citep{Chen1995} stimulations rapidly saturates.

\section{Discussion}
This paper proposes a simple model for the growth at high population density of plants competing for light. 
The model is based on the consideration that the plant growth is controlled by the auxin synthetic pathway, which is regulated by two signals, an internal one related to the hormone ethylene, and an environmental one, due to the mutual shading.
Moreover, the auxin signal is converted into biomass only after a time lag, $\alpha$, that varies on time. This memory effect forbids modelling the system markovianly at the natural time scale. 
In fact, a single plant has the predisposition to avoid the shade experienced at time $t$ only by shooting at time $t+\alpha$. At the system level, this can be modelled by maximizing the global shade avoidance potential~(\ref{hamcl}).

Consequently, this model cannot be considered to belong to any of the three frameworks of spatial community models drafted by \citet{Bolker2003}, namely the Interacting Particle Systems, the Stochastic Point Processes, and the Patch Models. In fact, even sharing some features with IPSs, the model proposed here is non-markovian. Indeed, it can be considered closer to those models that describe in terms of energy-like function(al)s a broad range of life science phenomena, such as X chromosome inactivation \citep{Nicodemi2007}, epithelial cell polarization \citep{Veglio2009}, and eukaryotic chemotaxis \citep{FCG+08,GKL+08}. 

The present work shows that the simulated plant height distributions are bimodal and right skewed, as in real communities of plants at high population density.
On the contrary, if no delay is assumed between the auxin signal and the biomass production, i.e. if the time lag $\alpha$ is set to zero, simulations show that the distribution of plant heights is unimodal and cannot reproduce real distributions.

Extending the model to isolated plants, an analytical expression is obtained both for the growth dynamics~(\ref{heightSec}) and the growth speed~(\ref{solgammaSec}). They have been fitted to {\it Arabidopsis thaliana} first inflorescence length data and on first inflorescence internode extension rate data, respectively. Theoretical functions and experimental data show a good agreement. 

Moreover, the obtained growth dynamics~(\ref{heightSec}) is consistent with the biomass production function proposed by \citet{Cournede2008}. 

The bimodality in height distribution, the good match between experimental data and theoretical functions for isolated plants, and the consistency with an empirical model independently derived, can be considered the main results of this work. They hold as long as the time lag $\alpha$ is taken into account. If the model is forced to be markovian, $\alpha=0$, real distributions and dynamics can no longer be mimicked. 
This suggests that memory effects play a key role in the plant growth process and that they cannot be neglected.   

Interestingly, \citet{Hara1984} studies the dynamics of plant populations with a stochastic model based on the Kolmogorov forward equation, which needs markovianity, and observes that the Markov property applied to plant growth is reasonable as a first approximation and needs further investigation. In fact, it seems not to be a case that his model does not reproduce bimodality in plant height distributions.  

A further step in the comprehension of plant growth would be to experimentally check the effects of varying the time lag between the auxin production and its conversion in biomass (for example, along the lines of \citet{Tao2008}), both in high population density and isolated plants.   
Accordingly, from a theoretical point of view it would be interesting to study the bimodal-to-unimodal switch in height distribution due to the loss of markovianity, in particular in terms of statistical physics of phase transitions \citep{Huang1987,b-a}. 

\section*{Appendices}
\appendix
\section{Analysis of the Dynamics}
\label{app0}
\renewcommand{\theequation}{\ref{app0}\arabic{equation}}
\setcounter{equation}{0}  
To analyze the dynamics, it is worth switching from discrete to continuous time by substituting $t\rightarrow N^2\ t$. This yields (\ref{spin}) to read 
\ba\label{spi}
\dot{s}_{ij}(t)=\rho\ e^{-b N^2 t}\ \eta_{ij}(t),
\ea
where 
\ba\label{cnoise}
\eta_{ij}(t)\equiv \xi(t)\ \Theta \big[A(t)-A(t-\alpha)\big]
\ea 
is a random term. I will treat it as a gaussian non-zero-mean colored noise with time-independent mean, which I shall label as 
\ba\label{lambda}
\Omega_{ij}\equiv\langle\eta_{ij}(t)\rangle=\langle\xi(t)\ \Theta \big[A(t)-A(t-\alpha)\big]\rangle.
\ea 
Therefore, the average of (\ref{spi}) over different system's realizations is simply
\ba
\label{ode1}
\avg{\dot{s}_{ij}(t)}= \rho\ \Omega_{ij}\ e^{-bN^2t},
\ea 
with the straightforward solution
\ba
\label{sol}
\avg{s_{ij}(t)}
=\frac{\rho\ \Omega_{ij}}{b\ N^2}\big(1-e^{-bN^2t}\big),
\ea 
which has the following steady state,
\ba
\label{u}
\langle s_{ij}\rangle_{st}\equiv\lim_{t\rightarrow\infty}\avg{s_{ij}(t)}=\frac{\rho\ \Omega_{ij}}{b\ N^2}.
\ea
The dynamics of an isolated plant height over time $q(t)$ may be defined as 
\ba\label{new}
q(t)\equiv\langle s(t)\rangle\big\arrowvert_{t=t-\alpha(t)}\ ,
\ea 
where $\langle\dot{s}(t)\rangle$ is the same as in~(\ref{ode1}), but without position indexes because it is referred to an isolated plant. It is calculated at time $t-\alpha(t)$ to reflect that auxin is not immediately converted into biomass, as discussed in Section~\ref{model}. (\ref{new}) is the same as (\ref{qq}) except for the average, which is intended to reduce fluctuations due to the isolated plant condition. The shade avoidance potential is ideally evaluated in the condition $J=0$, i.e. no interaction, and $N=1$. 

Using (\ref{alpha}), (\ref{sol}) and (\ref{new}), ${q}(t)$ reads 
\ba\label{height}
{q}(t)&=&\langle {s}(t)\rangle \big\arrowvert_{t=t-\alpha(t)}\\
\nonumber&=&{\rho\ \Omega}\ b^{-1}\ (1-e^{-b(t-m t/(K+t))}),
\ea 
and
\ba\label{solgamma}
\dot{q}(t)=\rho\ \Omega\ \Big(1-\frac{m\ K}{(K+t)^2}\Big)\ e^{-b(t-m t/(K+t))}.
\ea
The fact that $\alpha$ tends to the finite value $m$ implies that at the equilibrium the delay gets negligible, letting $q(t)=\langle s(t)\rangle$ (\ref{sol}) and $\dot{q}(t)=\langle \dot{s}(t)\rangle$ (\ref{ode1}). This is consistent with (\ref{qqq}), that holds after the equilibrium.   

\section{Analysis of the Parameters}
\label{app1}
\renewcommand{\theequation}{\ref{app1}\arabic{equation}}
\setcounter{equation}{0}  
In order to explain the model dependence on the parameters, it is worth evaluating analytically the noise mean $\Omega_{ij}$ introduced in~(\ref{lambda}). Its definition is general and valid for the whole system of plants.

(\ref{spi}) is a Langevin-like equation with a non-zero-mean colored noise, no drift term and time dependent diffusion term. It is possible to perform some transformations to obtain a more standard Langevin-like equation. Defining
\ba\label{x}
x_{ij}(t)\equiv s_{ij}(t)e^{b N^2 t},
\ea 
and inserting (\ref{spi}) in the time derivative of (\ref{x}), I obtain 
\ba\label{x.}
\dot{x}_{ij}(t)=F_{ij}(x)+\rho\ \tilde{\eta}_{ij}(t),
\ea
where
\ba
F_{ij}(x)&\equiv& b N^2 x_{ij}(t)+\rho \Omega_{ij},\\
\tilde{\eta}_{ij}(t)&\equiv& \eta_{ij}(t)-\Omega_{ij}.
\ea
Consider that $x(t)$ and $s(t)$ have, at the steady state, the same \e{pdf}.

(\ref{x.}) is a Langevin-like equation with time independent drift term, constant diffusion term and zero-mean colored noise.\\
Assuming that such a noise is a Ornstein-Uhlenbeck process \citep{Kam07}, i.e.
\ba
\frac{d \tilde{\eta}_{ij}(t)}{dt}=-\frac{1}{\mu_{ij}}\ \tilde{\eta}_{ij}(t)+\zeta(t),
\ea
where $\mu_{ij}$ is the correlation time and $\zeta(t)$ is a white noise with zero mean and delta time-correlation, it is possible to write the Fokker-Planck equation corresponding to (\ref{x.}), i.e.
\ba\label{o-u}
\frac{\partial p_{ij}(x,\eta,t)}{\partial t}=&-&b N^2+\gamma p_{ij}(x,\eta,t)+\\
&+&\gamma \eta\ \frac{\partial p_{ij}(x,\eta,t)}{\partial \eta}+\frac{1}{2}\ \frac{\partial^2 p_{ij}(x,\eta,t)}{\partial \eta^2}.\nonumber
\ea 
Even at the steady state, (\ref{o-u}) is not easy to be analytically solved.
However, for a huge set of parameters (\ref{x.}) lies in the validity range of the Unified Colored Noise Approximation, UCNA \citep{bow,raf,rain}. In fact, to apply the UCNA it is necessary \citep{raf} 
that
\ba\label{r1}
x_{ij}\gg \frac{\mu_{ij}}{b}.
\ea
According to (\ref{x}) and to the fact that $s_{ij}\sim 1$, constraint (\ref{r1}) corresponds to 
\ba
t\gg \frac{\mbox{log}(\mu_{ij}/b)}{b N^2}\equiv t_{u}.
\ea
Therefore, since equilibrium time $1/b\gg t_{u}$, the UCNA can be applied for the parameter set used in the reported numerical simulations. Applying UCNA, the following effective Langevin equation holds 
\ba\label{eff}
\dot{x}_{ij}(t)\simeq\sigma_{ij} F_{ij}(x)+\rho\ \sigma_{ij}^2\zeta(t),
\ea
where $\zeta(t)$ is again a white noise with zero mean and delta time-correlation, while I defined $\sigma_{ij}$ as 
\ba
\sigma_{ij}\equiv \frac{1}{1-\mu_{ij} b N^2}.
\ea
I calculated the correlation time $\mu_{ij}$ by averaging the numerical values of $\alpha$ over all the lattice nodes and over different realizations of the system. As an example, labelling the site at the center of the lattice as $cc$, the sites at the center of the lattice borders as $bb$ and the sites at the corners as $aa$, I obtained, in continuous time units, $\mu_{cc}\simeq54$, $\mu_{bb}\simeq50$ and $\mu_{aa}\simeq47$.

\begin{figure}
\centering
\includegraphics[scale=.45]{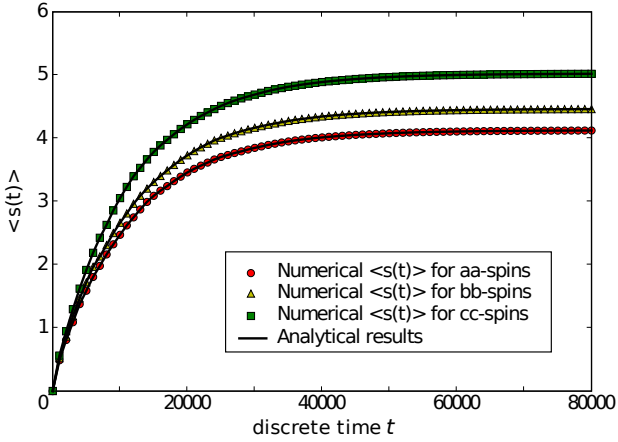}
\caption{Eq.~(\ref{sol}) (lines) and numerical results from the simulation of the dynamics~(\ref{spin}-\ref{ham}) (symbols). Arbitrary units.}
\label{s_vs_t}
\end{figure}

In passing, notice that the drift term of (\ref{eff}), $\sigma_{ij} F_{ij}(x)$, is linear in $x_{ij}$; therefore this model lies in a class of universality different from the principal growth models\footnote{In particular, in the Kardar-Parisi-Zhang model \citep{kpz} 
the drift term is equal to $\nu \nabla^2 x_{ij}+\frac{\mu}{2}(\nabla x_{ij})^2$, where $\nu$ and $\mu$ are characteristic parameters of that model.} \citep{b-a}.
Then, it is possible to write the Fokker-Planck equation corresponding to (\ref{eff}),
\ba
\frac{\partial p_{ij}(s,t)}{\partial t}=&-&\sigma_{ij}\frac{\partial\big( F_{ij}(s)  p_{ij}(s,t)\big)}{\partial s}+{}\nonumber\\
&+&\frac{1}{2}\rho^2\sigma_{ij}^4\frac{\partial^2 p_{ij}(s,t)}{\partial s^2},
\ea 
and consequently to calculate the probability distribution function at the steady state $p_{ij}(s)\equiv \lim_{t\rightarrow\infty}p_{ij}(s,t)$ \citep{Kam07}
\ba
p_{ij}(s)=\frac{\mbox{exp}[-\rho^{-2}\sigma_{ij}^{-3}(bN^2s^2+2\rho\Omega_{ij}s)]}{\int_0^1dy\ \mbox{exp}[-\rho^{-2}\sigma_{ij}^{-3}(bN^2y^2+2\rho\Omega_{ij}y)]}.
\ea
Therefore,
\ba\label{s}
\langle s_{ij}\rangle_{st}\simeq\int_{0}^{\infty}ds\ s\ p_{ij}(s)\equiv g(\Omega_{ij},\dots),
\ea
where the last term is to express that the analytical solution of the integral in (\ref{s}), that I do not report for reasons of shortness, is a function of the noise mean $\Omega_{ij}$. Comparing (\ref{s}) to (\ref{u}) the following implicit equation in $\Omega_{ij}$ holds
\ba\label{l}
\rho \Omega_{ij}\simeq bN^2g(\Omega_{ij},\dots),
\ea
which has solutions, for plants in different position of the lattice labelled as above, \mbox{$\Omega_{cc}\simeq.23$}, \mbox{$\Omega_{bb}\simeq.20$} and \mbox{$\Omega_{aa}\simeq.18$}.
In Fig. \ref{s_vs_t} analytical solution (\ref{sol}) and numerical results are plotted.
\section*{Acknowledgements}
The author wishes to thank Federico Bussolino, Michele Caselle, Antonio Celani, and Marta Lucchini for useful discussions and comments, and Guido Serini for carefully reading the manuscript.

\end{document}